\documentclass[12pt]{article}

\usepackage{newtxtext,newtxmath}     % Times-family text + matching math
\usepackage[T1]{fontenc}
\usepackage{microtype}               % proper protrusion + expansion;
\usepackage[letterpaper,margin=1.1in]{geometry}

\usepackage{setspace}
\usepackage{booktabs}
\usepackage{graphicx}
\usepackage{url}
\usepackage{comment}

\usepackage{titlesec}
\titleformat{\section}
  {\normalfont\Large\bfseries}{\thesection}{0.6em}{}
\titleformat{\subsection}
  {\normalfont\large\bfseries}{\thesubsection}{0.5em}{}
\titleformat{\subsubsection}
  {\normalfont\normalsize\bfseries}{\thesubsubsection}{0.5em}{}
\titlespacing*{\section}      {0pt}{1.6ex plus .4ex minus .2ex}{0.9ex plus .2ex}
\titlespacing*{\subsection}   {0pt}{1.3ex plus .3ex minus .2ex}{0.7ex plus .2ex}
\titlespacing*{\subsubsection}{0pt}{1.0ex plus .3ex minus .2ex}{0.5ex plus .2ex}

\usepackage[font=small,labelfont=bf,labelsep=period]{caption}

\usepackage{fancyhdr}
\usepackage[colorlinks=true,
            linkcolor=black,
            citecolor={blue!45!black},
            urlcolor={blue!55!black}]{hyperref}

\usepackage{scicite}

\makeatletter
\renewcommand{\fnum@figure}{\textbf{Figure \thefigure}}
\renewcommand{\fnum@table}{\textbf{Table \thetable}}
\makeatother

\def\scititle{Nothing Deceives Like Success: Social Learning and the
              Illusion of Understanding in Science}
\title{\bfseries\boldmath \scititle}

\author{%
  Avery W. Louis\,$^{1\ast}$\quad Marina Dubova\,$^{2}$ \\[0.6ex]
  \normalsize $^{1}$Symbolic Systems, Stanford University, Stanford, CA 94305, USA \\
  \normalsize $^{2}$Santa Fe Institute, Santa Fe, NM 87501, USA \\[0.4ex]
  \normalsize $^{\ast}$Corresponding author. Email: \texttt{averylou@stanford.edu}
}
\date{}

\begin{document}
\maketitle
\thispagestyle{fancy}                % header on the first page too

\begin{abstract}
\noindent
Success-driven social learning, in which individuals preferentially adopt the
ideas and methods that appear most successful, is a foundational principle of
collective behavior across systems ranging from ant colonies to scientific
communities. But science is a particular kind of collective search---one in
which the quality of an explanation is itself difficult to assess. Is success
bias adaptive in this setting? In agent-based simulations of collective theory
building, we find that it is not. Scientists in our model systematically
overestimate the quality of their own theories, creating an \textit{illusion
of understanding}: a persistent gap between perceived and actual performance.
Success bias amplifies this illusion; communities that favor apparently
successful theories explore a narrower range of possibilities, efficiently
filtering out poor explanations but failing to discover better ones. This
effect intensifies with problem complexity, as scientists in more complex
environments become increasingly unable to assess how well their theories
actually perform. Most strikingly, when agents optimize their social behavior
to maximize the perceived success of their theories, they paradoxically
undermine their actual performance, and produce levels of inequality that
mirror those found in real scientific communities.\footnote{All code, data,
and simulation outputs are available at
\url{https://osf.io/u3xc9/?view_only=309cb49465554a869932147b1ab74b50}.}
\end{abstract}

\noindent

Science is a fundamentally social enterprise. If we had no means of building on the work of others, we couldn't possibly hope to build thinking machines, create life-saving vaccines, or go to the moon and beyond. But the cumulative nature of science introduces a critical question: \textit{how} should scientists learn from one another? Every scientist must decide whose work to build on, which ideas to adopt or abandon, and where to focus their limited attention---decisions that collectively shape the trajectory of research. Of course, there are many ways to navigate these decisions, and a substantial body of work across social psychology, collective behavior, and cultural evolution has explored how individuals do so \cite{Boyd_1988, Laland_2004, Goldstone_2013, Barkoczi_2016}. Among the possible approaches, \textit{success-biased} learning, wherein scientists preferentially adopt and build on the ideas that appear ``best", seems to be a neatly rational approach \cite{Rendell_2010, Baldini_2012, wu2025adaptive}. If, for example, you were a cosmologist hoping to explain dark matter, a very natural first step would be to ask which theories currently seem most promising and work on one of those. It is easy to see why: promising theories are promising for reasons (they might fit more data, have fewer free parameters, or connect to other successful frameworks) so building on existing success seems like a rational use of limited time and attention. By doing so, communities can, in principle, concentrate their collective efforts on the most promising solutions, weeding out dead ends.

Success bias also aligns with one of the most powerful ideas in both nature and engineering---that systems improve by selecting for what works. It is the logic of natural selection, of ant colonies reinforcing productive foraging trails, and of gradient descent in optimization. Much contemporary science policy pushes in precisely this direction; hiring and promotion increasingly rely on quantitative metrics of success such as citation counts and h-indices \cite{Hirsch_2005}, funding agencies favor proposals with demonstrated track records of results, and there is a broader institutional push to make science more ``accountable" through measurable outputs \cite{REF_2029, Hicks_2015}. The implicit assumption is that science would improve if it were more systematically success-biased.

But this assumption rests on two premises: that we can reliably identify which scientific contributions are successful, and that following the most successful leads to genuine discovery. Recent computational work challenges the second premise, showing that success-biased strategies can impede innovation \cite{Wu_2024}. Here, we challenge both. What makes a scientific theory ``good"? The straightforward answer, that good theories accurately describe reality, obscures a fundamental epistemic constraint: scientists never have direct access to how well their theories map onto the world. Instead, the quality of theories scientists produce can only be estimated indirectly, through proxies such as predictive accuracy, empirical fit over collected data, parsimony, coherence, and others. The history of science is marked by episodes in which this process of evaluation has gone awry. Scientists routinely work with theories that appear adequate, or even compelling, only to later discover they systematically fail to account for observations at hand \cite{Ioannidis_2005, Langmuir_1989, Potochnik_2020}. This happened in the decades-long focus on amyloid plaques as the primary cause of Alzheimer's disease, which resulted in a string of failed clinical trials \cite{kepp2023amyloid}. Occasionally, entire research programs revolve around phenomena that are later decided not to exist at all, as demonstrated by the case of N-rays, where dozens of physicists reported detecting a form of radiation that was ultimately shown to be illusory \cite{Nye_1980, Brewer_2012}. These episodes reveal an important limitation of the scientific enterprise: because all theories are provisional attempts to capture aspects of an incompletely observed world, it may be genuinely difficult, at any given moment, to assess how good they actually are \cite{McBride_2012, Tetlock_2017}. This results in a potential divergence between the scientific community's in-the-moment judgments of theoretical quality and the longer-term performance of those theories, called an \textit{illusion of understanding} \cite{Messeri_2024}. If such illusions are routine rather than exceptional, they carry significant implications for social learning in science, as the very signal that learners rely on may be systematically misleading.

Here, we investigate the consequences of success-biased social learning for collective theory development using 2,301 agent-based simulations. Our model captures key features of scientific practice: scientists actively sample data from their environment, construct compressed representations (theories) that explain their observations, and engage in social learning by sharing data and selectively adopting theoretical components from colleagues. We measure both \textit{perceived} theory success (how well each scientist's theory fits the data the scientific community has access to) and \textit{actual} success (how well theories account for the held-out ground truth), allowing us to directly measure illusions of understanding and track their effects.

We find that although illusions of understanding exist regardless of the social learning strategies employed, success bias amplifies them. When scientists in the model favor seemingly successful theories by their peers, they explore a narrower range of possibilities, efficiently discarding poor explanations but also missing opportunities to discover better ones. Moreover, social learners confronting more complex problems are especially prone to overestimating the success of their theories compared to those facing simpler tasks. Finally, when learners adjust their social strategies to maximize \textit{perceived} success, they paradoxically worsen \textit{actual} performance.

\section{Materials and Methods}
We extend an agent-based model of scientific inquiry to investigate how the structure of a learning problem and scientists' social learning strategies---in particular, the tendency to preferentially adopt theories that appear more successful---jointly shape the quality of developed theories and the social organization of scientific communities \cite{Dubova_2026}. Each agent in the model represents an individual scientist who conducts experiments and builds a theory of their environment based on both their own results and those of others. This framework allows us to explore how systematic misalignment between perceived and actual success can arise as a natural consequence of learning in complex environments, and how different social preferences shape agents’ ability to learn about the world.

The motivation for such a model is twofold. First, simulation allows us to systematically explore a range of strategies that are considered rational yet cannot be investigated in isolation for scientists in the real world. In real-world science, social learning is entangled with prestige, status, and institutional affiliation, making it difficult to isolate the effect of perceived theoretical success alone \cite{Merton_1968, Perc_2014, Clauset_2015}. Our model strips away these confounds, allowing agents to be influenced solely by the apparent performance of theories rather than by the reputation of their holders and other factors. This allows us to ask whether and when success-biased social learning is useful for scientific progress. Second, and perhaps more crucially, simulations  let us create an epistemic ground truth. In the real world scientists judge theories based on data they happen to have collected or other proxies of epistemic success. In contrast, a simulated world lets us define the underlying data-generating process explicitly. This enables us to evaluate not just whether or not scientists in the model reach a consensus, but whether that consensus aligns with the actual structure of the simulated world they set out to investigate.

\begin{figure}
    \centering
    \includegraphics[width=1\linewidth]{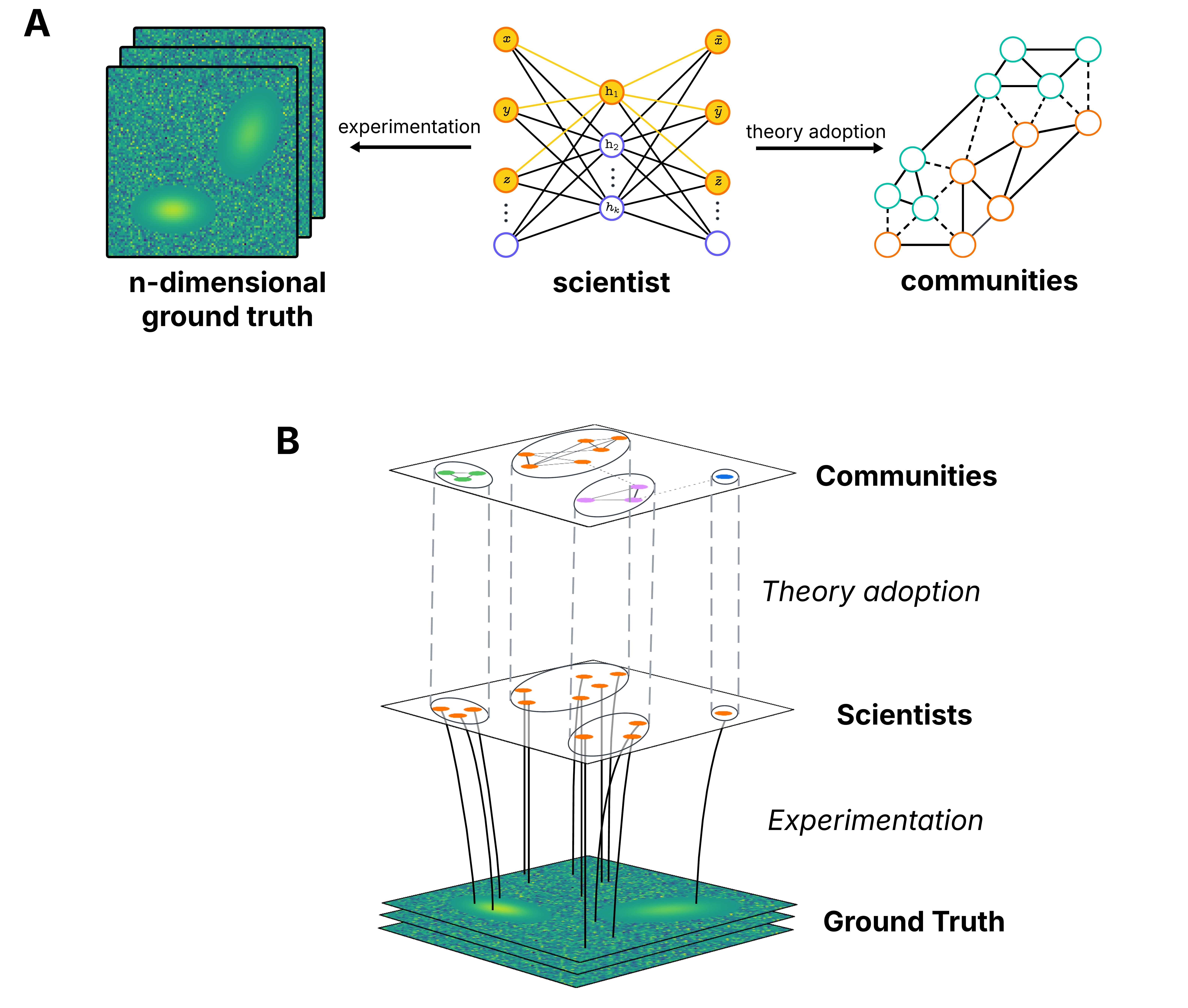}
    \caption{\textbf{Simulation overview.} \textbf{(A)} The ground truth consists of Gaussian distributions embedded in uniform noise. Scientists conduct experiments by querying specific configurations of variables and observing the returned values, then build theories---shallow autoencoders that learn compressed representations of the patterns in their data. Scientists evaluate and exchange theories via social learning, sometimes adopting components of another's theory based on its apparent success. Over time, these adoption dynamics give rise to emergent community structure.  \textbf{(B)} a representation of the model across three levels: the ground truth (environment), scientists (individuals), and communities (social).}
    \label{fig:sim}
\end{figure}

\subsection{Ground Truth}\label{sec:GT}

The ground truth represents a ``world” or ``phenomenon" which agents aim to learn about. This world is not uniformly structured, nor fully random---it contains both interpretable regularities and noise, reflecting the view that natural systems possess discoverable structure often only at particular scales\cite{Wimsatt_2021, Simon_1962, Cartwright_1999, Mitchell_2012}. Our goal is to investigate whether scientists in the model can, over time, distinguish signal from noise to form reliable theories of the signal based on the results of their experiments.

More precisely, the ground truth is implemented as an $n$-dimensional hypercube. Within this space, we place $n$-dimensional Gaussian distributions, each of which represents a locally structured learnable region. These Gaussians have randomly initialized means and covariance matrices, and are bounded. The sparsity of the ground truth is controlled by varying the number and spread of these Gaussians (see Table \ref{tab:parameters} for exact parameter ranges). In effect, these Gaussians define ``causal patches" within the world; regions where one variable is systematically related to others, and where patterns can be learned.

We operationalize complexity via the number of dimensions in the ground truth. As dimensionality increases, agents have to consider more variables simultaneously, and the potential for nonlinear interactions or synergistic effects among variables grows \cite{Simon_1962}. This makes pattern detection and generalization considerably more difficult, especially given finite data and theory-building constraints.

\subsection{Scientists}
In this model, agents represent individual scientists. Their task is to conduct experiments on the ground truth and develop theories to account for the results.

\subsubsection{Experimentation}
Scientists interact with the ground truth by sampling specific configurations of variables. Each sample returns a single observed value. If the configuration falls within the domain of a Gaussian, the returned value is governed by that Gaussian; otherwise, it is drawn from uniform noise.

\subsubsection{Theory Building}
Each scientist's theory is a shallow autoencoder: a simple neural network with a single hidden layer that predicts observed values from a set of input variables and their values. The hidden layer has a fixed width, which we call the explanation capacity, and it determines how much structure the scientist's theory can encode. A higher capacity allows the agent to represent more complex or nuanced patterns; lower capacity forces simpler, more compressed representations. Scientists' theories begin with randomly initialized weights and are fit to experimental results by minimizing the autoencoder's reconstruction error.

\subsubsection{Theory Scope}
Scientists can also specialize, focusing their experimentation and theory on a particular region of the ground truth (for example, some subset of variables or some subset of ranges within those variables). We model this through each scientist's \textit{scope}---an internal estimate of where their theory is relevant. This estimate is constructed via a kernel density estimate over past samples, weighted by inverse prediction error. The density estimate captures \textit{where} in the ground truth a scientist has experimented, and the weighting boosts regions where the theory performs well.

A \texttt{bandwidth} parameter controls how tightly or broadly a scientist's scope is drawn, governing the degree of specialization. Because scientists preferentially draw new samples from within their scope, bandwidth also controls experimentation strategy. Low-bandwidth scientists have a more confirmatory experimentation strategy, sampling near where their theory has performed well, while high-bandwidth scientists are more exploratory.

\subsection{Social Learning}
Just as working scientists do, agents in the model engage in social learning by sharing experimental data and adopting one another's theories. Real scientists employ a mixture of social learning strategies, whose effectiveness varies by context \cite{Laland_2004}. In our model, we focus on two: \emph{success-biased} social learning, in which agents preferentially interact with those whose theories appear most successful, and \emph{community-biased} social learning, in which agents preferentially interact with members of their own community. These biases govern two key processes: data exchange and theory adoption, and in turn give rise to emergent community structure.
\subsubsection{Data Exchange}
Every round of the simulation, agents share data. Two agents are selected at random (with a preference towards agents from the same community, determined by the degree of community bias) to exchange data from their most recent experiment with one another.
\subsubsection{Theory Evaluation}
One of the central focuses of our model is finding out whether or not success-biased social learning can produce illusions of understanding, where agents judge their theories to be performing well, despite them not accurately describing the ground truth. To capture this, we evaluate theories in two different ways: \textit{actual performance} measures how well theories capture the entire held-out ground truth, and is therefore inaccessible to agents, while \textit{perceived performance} measures how well theories capture the data that agents have collected, and is therefore the kind of evaluation agents have direct access to. 

\paragraph{Actual theory performance.} We measure a theory's actual performance by evaluating its predictions against a sample from the ground truth. We account for the fact that agents might specialize in different regions of the ground truth by sampling 1,000 points from the true data-generating process (see Section~\ref{sec:GT}) and assigning each point to the community of agents whose average sampling location is nearest\footnote{Using a Voronoi partition of the variable space by community}. Each agent's theory is then evaluated only on the points assigned to their community, using the mean absolute percentage error of their neural network's predictions versus the ground truth. This ensures that scientists' theories are only evaluated on the parts of the ground truth they are designed to explain, rather than penalizing specialized theories for not making good predictions outside their scope. Agents \textit{don't} have access to this kind of evaluation as it measures performance directly against the ground truth. Instead, they use perceived performance.

\paragraph{Perceived theory performance.} Perceived performance is the evaluation signal actually available to agents within the simulation. Each agent computes the mean absolute error of its own neural-network predictions against its personally collected data (the same observations it used to train its theory). This self-evaluated loss is the basis on which agents compare theories during social learning: when deciding whether or not to adopt a neighbor's theory, agents can only judge theory quality relative to data the community has already gathered. A theory, then, could score well on perceived performance despite poor actual performance if the agent's sampling has been concentrated in a region the theory happens to fit, or if the agent has not yet found parts of the environment where their theory fails.

\subsubsection{Theory Adoption} \label{sec:theory eval}
Scientific practice often involves working on or borrowing from existing theories. In our model, agents adopt components of, or entire theories from other agents. At each simulation round, a subset of scientists undergo theory adoption, governed by two parameters: \textit{success bias}, which determines how strongly agents prefer theories that appear more successful, and \textit{community bias}, which determines how strongly agents prefer theories from within their own community. When both are low, adoption is approximately random.

\paragraph{Step 1: Selecting an Adopter.} First, all the agents are ranked according to the performance (MAE) of their current theory on the data the community has collected. Agents with poorer-performing theories are more likely to be selected as adopters. Specifically, the probability of selecting scientist $i$ as an adopter is given by a temperature-scaled softmax over losses:
\begin{equation}
    p_i^{\text{adopt}} = \frac{\exp\left(\frac{\ell_i}{T}\right)}{\sum_{j=1}^{n} \exp\left(\frac{\ell_j}{T}\right)}
\end{equation}

\noindent where $\ell_i$ is the loss of agent $i$'s theory and $T$ is the temperature parameter. Because higher loss gives a higher probability, agents with poorer-performing theories are more likely to be selected. Low temperatures amplify this bias (strong selection against poorly performing theories), while high temperatures flatten the distribution (more randomness).

\paragraph{Step 2: Selecting an Adoptee.} Once the adopter is chosen, an adoptee is selected (an agent whose theory will be copied from). To favor agents with \textit{better} theories, selection is based on scores $q_i = \max_j(\ell_j) - \ell_i$, so that lower-loss agents receive higher scores. The probability that agent $i$ is selected as the adoptee is proportional to both their score and community alignment:

\begin{equation}
    p_i = \frac{\exp\left(\frac{q_i}{T}\right) \cdot b_i}{\sum_{k=1}^{n} \exp\left(\frac{q_k}{T}\right) \cdot b_k}
\end{equation}

\noindent where $b_i = \textit{community bias}$ if agent $i$ is in the same community as the adopter, and $b_i = 1 - \textit{community bias}$ otherwise. Together these parameters modulate agents' preferences over whom to socially learn from, but when both success and community biases are low, copying is random.

\begin{figure}
    \centering
    \includegraphics[width=1\linewidth]{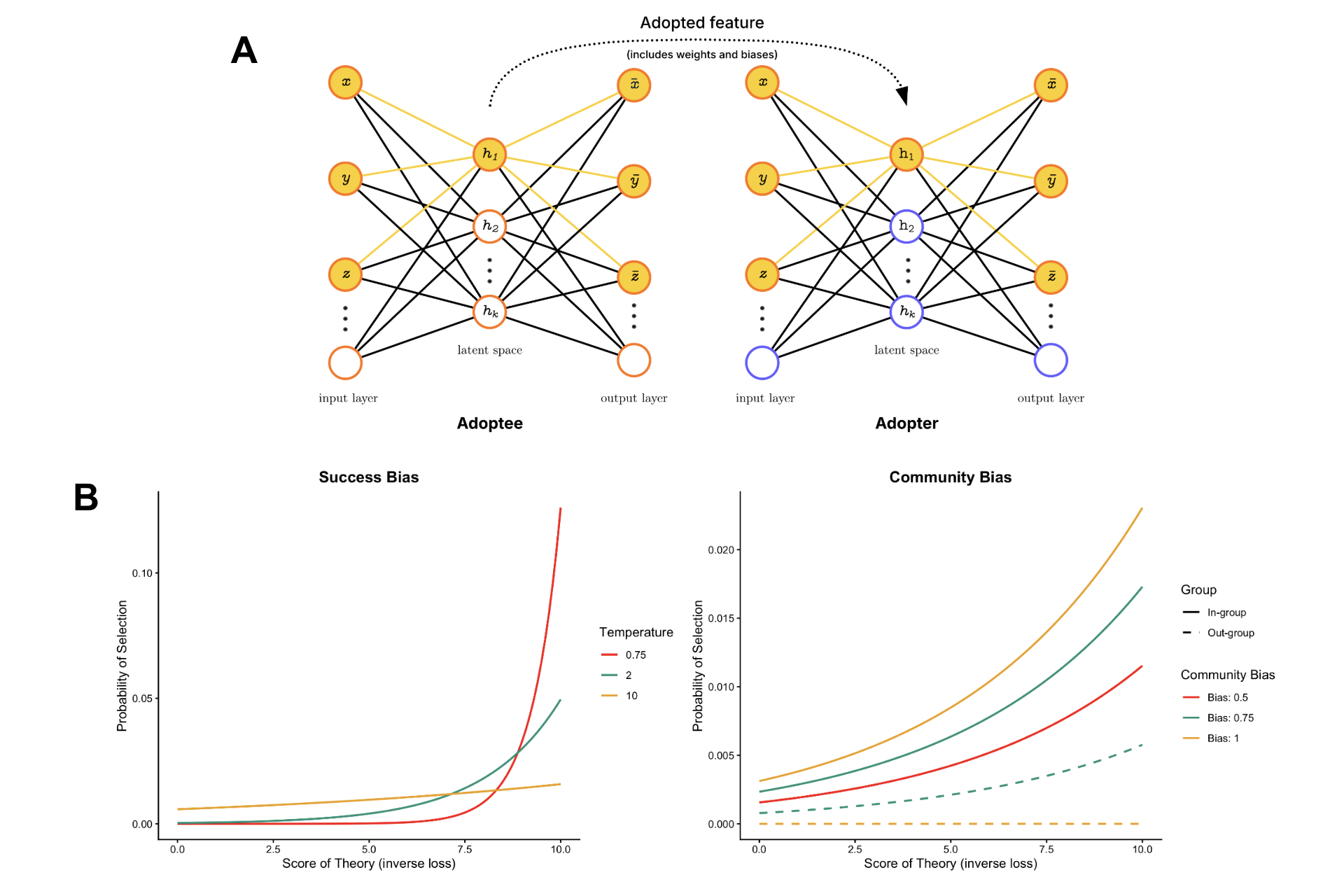}
    \caption{\textbf{Aspects of social learning.} \textbf{(A)} Theory adoption between two autoencoders. The adopter copies one or more entire features from the adoptee's neural-network over to their own network. \textbf{(B)} The effect of success bias and community bias on theory adoption. Success bias (left) controls how strongly selection favors high-scoring theories: low temperature concentrates probability on top performers, while high temperature approaches random selection. Community bias (right) modulates selection by group membership. Solid and dashed lines of the same color show in-group and out-group selection probabilities respectively. At maximum bias (1.0), out-group agents have zero probability of selection; at minimum (0.5), group membership has no effect. Both biases operate jointly in the model, as shown by the combined effect on the right.}
    \label{fig:SB_CB}
\end{figure}

\paragraph{Step 3: Partial Theory Transfer.} The selected adopter then copies part of the adoptee’s theory by copying a number of \textit{features}---individual hidden units and their associated input and output weights---from the adoptee’s autoencoder. Before this happens, we align the two theories \cite{Kuhn_1955}. The number of transferred features is proportional to the relative performance of the adoptee’s theory, and bounded above by a fraction (max exchange) of the adopter’s total hidden units. This ensures that high-performing theories exert greater influence.\\

Each hidden unit in the autoencoder learns to detect a recurring pattern in the agent's data (i.e. a regularity about how some subset of variables covary). The incoming and outgoing weights of a unit together form a basis function in the agent's approximation of the world \cite{Bengio_2014}, so copying a single unit as shown in ~Figure~\ref{fig:SB_CB} Panel A transfers one interpretable regularity, roughly analogous to adopting a single equation from a colleague's model or importing a construct while keeping the rest of one's framework intact.

\subsubsection{Community Formation}

As agents interact and share information over time, they begin to form epistemic communities. These communities reflect persistent patterns of influence, and they provide a social structure through which theories propagate, given the community bias outlined above.

At the start of the simulation, agents are unconnected individuals. During each round of the simulation, whenever an agent adopts all or part of another agent’s theory, a directed edge is added from the source (adoptee) to the target (adopter). To identify community structure, we apply the Louvain method for community detection to the changing adoption graph every five rounds \cite{Blondel_2008}.

\subsubsection{Other evaluative metrics}\label{sec:other_eval}
We also measure some features of scientists' theory-making abilities as a population. These are metrics that the scientists themselves \textit{don't} have access to.

\paragraph{Theory Diversity.} Because scientists in our model specialize in different regions of the ground truth, we are less interested in whether they collectively cover the entire environment, which would necessitate having a diverse set of theories, than in whether scientists working on the same phenomena maintain distinct explanations for it. We therefore measure theory diversity within communities for both of the diversity measures that follow.

The first operates in \textit{weight-space}: within each community, we compute the average pairwise distance between members' theory parameters. Because neural networks that produce the same outputs can have permuted hidden units, we first align each pair of theories using the Hungarian algorithm to find the best matching of hidden units before computing the euclidean distance between their weights \cite{Kuhn_1955}. Higher values indicate that scientists from the same communities have more diverse theories, and lower values show a convergence toward similar ones.

However, weight-space distances can be a poor proxy for functional differences between the actual neural networks themselves because networks with very different weights can compute similar functions. Further, the Euclidean distance between theories can be confounded by explanation capacity which increases the number of hidden neurons to compare, causing the total weight-space to grow. We therefore also measure diversity in \textit{function-space}. To do so, we sample 500 points from the ground truth and collect each scientist's predictions at every point. We then assign evaluation points to communities using a Voronoi partition based on each community's average sampling location, and find the mean pairwise Euclidean distance between members' \textit{predictions} on their assigned points. The overall measure averages these within-community values. Higher values show divergent predictions among scientists studying the same phenomena, whereas lower values show convergence.

\paragraph{Evaluative Accuracy (CrPC).} We measure whether the scientific community can accurately identify its best members by computing the Pearson correlation between each scientist's influence in the adoption network (measured by Katz centrality) and their actual performance (section \ref{sec:theory eval}) relative to the community average. This is a Centrality and relative Performance Correlation Coefficient (CrPC), and represents scientists' evaluative accuracy. A high positive correlation means influential scientists are genuinely good, so the community's social structure accurately tracks quality. A low or negative correlation indicates that popularity is decoupled from, and sometimes even inversely related to, actual performance. 

\subsection{Bayesian Optimization}
To investigate which social learning strategies agents would adopt if they 
were rational optimizers of their own perceived theory success, we conducted an  inverse design experiment using Bayesian optimization \cite{Shahriari_2016}. We treated four social parameters of the model (success bias, community bias, degree of sociality, and explanation capacity) as tunable variables, with the objective of finding the configuration that maximizes the median perceived success of the scientific community as a whole.

We repeated this procedure independently across six environments varying 
in problem complexity (simple, medium, complex) and sparsity (dense, 
sparse). Each optimization run had two phases: an initial Sobol 
sampling phase of 20 steps to broadly explore the parameter space 
\cite{Sobol_1967}, followed by 20 steps of Gaussian process-guided 
Bayesian optimization. Because other parameters of the model were set randomly and can introduce noise into the outcomes, each configuration was evaluated across 3 independent simulation replicates, with their median perceived success scores averaged. This gave a more stable estimate of each configuration's perceived success score.

\subsubsection{Empirical Citation Inequality}\label{sec:gini_methods}
To compare the inequality produced by our optimization procedure against real-world science, we estimated Gini coefficients from citation data from Scopus. For each of five fields (Astrophysics, Cognitive Science, Genetics, Physical Chemistry, and Physics), we randomly sampled 20,000 articles published between 2000 and 2020. We used fractional counting \cite{Nielsen_2021}: dividing each article's citation count by its number of authors, then aggregating at the level of authors to count cumulative fractional citations per author.

\subsection{Procedure}\label{sec:proc}
In each simulation, $N$ of the randomly initialized agents learn about the ground truth, and by the end of the simulation 2500 experiments are conducted in total. This proceeds over 300 `rounds', where in each round a number of agents may interact through data sharing or theory adoption. This number is governed by agent sociality. We also measure the success of agents' theories against their collected datapoints (perceived success) as well as the ground truth (actual success) (see section~\ref{sec:theory eval}), the diversity of theories, and properties of the adoption graph. Overall, we ran the simulation 2301 times with random initialization of parameters in each run (see Table~\ref{tab:parameters}).

To explore the behavior of our model across a wide range of plausible scientific environments, we adopted a radical randomization strategy \cite{Baribault_2018}. Every core parameter of the simulation (including the number of scientists, explanation capacity, success and community bias, and scope bandwidth) is drawn randomly from a uniform distribution over a broad but reasonable range. This produces noisy results but ensures that any patterns we observe are robust across diverse configurations of agents and environment. 

\begin{table}[t!]
\centering
\caption{The ranges of parameter values for the simulations.}
\label{tab:parameters}
\begin{tabular}{lcc}
\toprule
\textbf{Parameter} & \textbf{Range} & \textbf{Type} \\
\midrule
\multicolumn{3}{l}{\textit{Environment}} \\
\quad Dimensions          & $[3, 8]$      & Discrete             \\
\quad Sparsity            & $[5, 75]$     & Continuous           \\
\midrule
\multicolumn{3}{l}{\textit{Social}} \\
\quad Explanation Capacity & $[3, 8]$     & Discrete             \\
\quad Bandwidth           & $[0, 2]$      & Continuous           \\
\quad Temperature (Success Bias) & $[0.05, 10]$ & Continuous (log scale) \\
\quad Community Bias      & $[0.5, 1]$      & Continuous           \\
\quad Max Exchange        & $[0.5, 1]$    & Continuous           \\
\midrule
\multicolumn{3}{l}{\textit{Simulation}} \\
\quad Number of Agents    & $[20, 200]$   & Discrete             \\
\quad Sociality    & $[0.1, 0.25]$ & Continuous           \\
\quad Number of Datapoints & $2500$       & Fixed                \\
\quad Rounds              & $300$         & Fixed                \\
\bottomrule
\end{tabular}
\end{table}

\section{Results}
\begin{figure}
    \centering
    \includegraphics[width=1\linewidth]{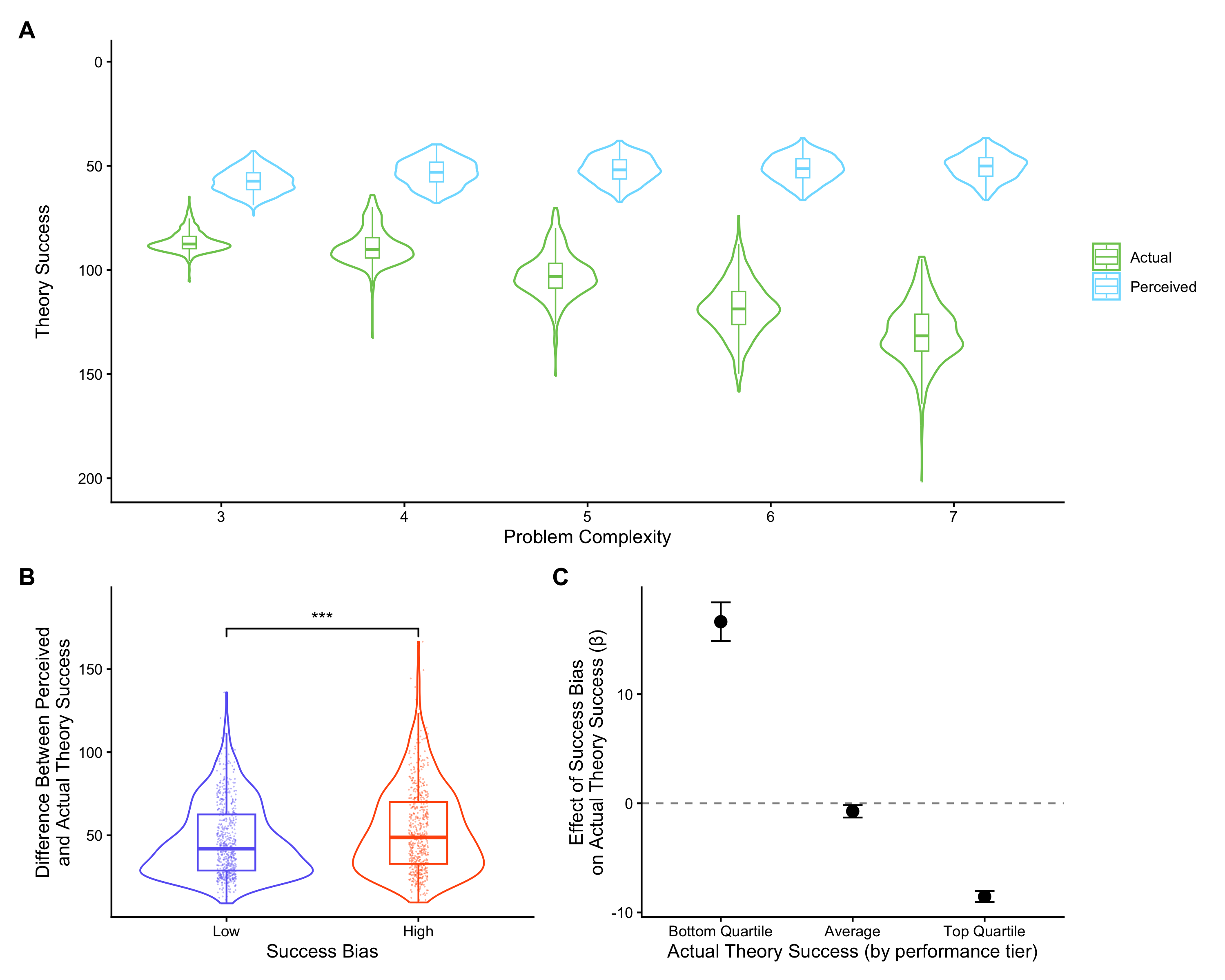}
    \caption{\textbf{Illusions of Understanding.} \textbf{(A)} Perceived theory success (teal) remains stable across levels of problem complexity, while actual theory success (green) declines, creating a growing gap between what agents believe about their theories and how those theories actually perform. Note: theory success is reversed, as it is measured as theory loss. \textbf{(B)} This gap between perceived and actual success is larger under high success bias (red; top quartile of success bias, $n=576$) than low (purple; bottom quartile of success bias, $n=576$), indicating that success bias exacerbates illusory confidence in theory quality (Mann-Whitney $U = 143,688,\; p < .001$). \textbf{(C)} Regression coefficients for the effect of success bias on actual theory success by performance tier. Success bias improves bottom-quartile theories but degrades top-quartile ones, with a negligible effect on the average.}
    \label{fig:illusions}
\end{figure}

We ran 2,301 simulations to examine how the illusion of understanding depends on the social learning strategies employed by scientific communities. We found that the gap between perceived and actual theory success arises systematically and widens substantially as problem complexity increases (given that problem complexity is ordinal, we use Spearman's rank correlation $\rho(2299) = .817, p < .001$), indicating that agents become increasingly unable to make accurate assessments of how well their theories perform in more complex environments. The gap was significantly greater than zero \textit{both} in the absence of social learning (Wilcoxon signed-rank $(N=268)$, $V=36,045,\; p<.001$) \textit{and} across all social learning conditions (Wilcoxon signed-rank $(N=2,301)$, $V=2,648,451, \; p < .001$), meaning the illusion of understanding stems from structural features of the learning environment and properties of the individual learning process, such as possibly overfitting theories (Figure \ref{fig:illusions}, panel A). However, the severity and consequences of such illusions of  understanding \textit{do} depend on the social learning strategies employed by the agents. 

We examine these effects next by looking at how success bias, or the tendency to learn preferentially from theories that appear more successful, affects social dynamics (\ref{sec:SB and social dynamics}), the success of theory development (\ref{sec:SB and theory dev}), the degree of diversity among theories pursued by the community (\ref{sec:SB and div}), agents' ability to evaluate the performance of a given theory (\ref{sec:CrPC}), and susceptibility to overfit theories (\ref{sec:overfitting}).

\subsection{Success Bias and Social Dynamics}\label{sec:SB and social dynamics}
\begin{figure}
    \centering
    \includegraphics[width=1\linewidth]{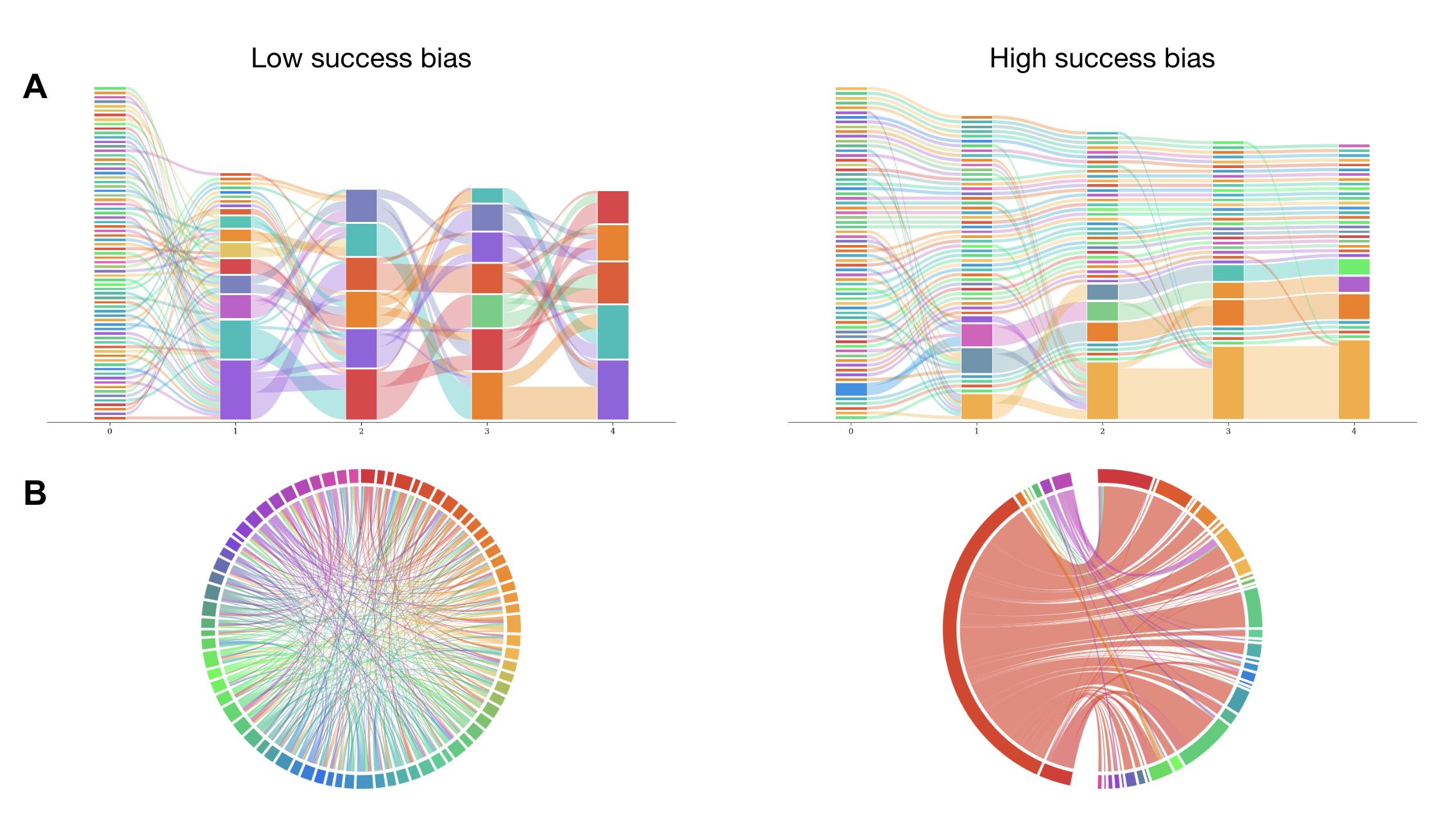}
    \caption{\textbf{Community structure and information flow under low and high success bias.} \textbf{(A)} Alluvial diagrams tracking community membership across simulation rounds for representative simulations (with 70-75 agents, and little to no community bias) under low (left) and high (right) success bias. Each vertical axis represents a time step in the simulation, with all agents beginning unaffiliated with any community. Colored streams trace individual agents' community affiliations over time. \textbf{(B)} Chord diagrams showing the direction and magnitude of theory exchange between agents at the end of the same two simulations. Any agents who never adopted other's theories are not included. Ribbon color indicates the adoptee (i.e., the theory being adopted, or information flow), and ribbon width indicates the volume of exchange.}
    \label{fig:alluv}
\end{figure}

Overall, success bias significantly reshapes the structure of information exchange between scientists. In our simulations, we found that success bias discourages heterogeneous information flow between agents, instead concentrating attention on the few most successful agents (see Figure \ref{fig:alluv}).

Figure \ref{fig:alluv}, panel A illustrates these differences using alluvial diagrams from single simulations that track scientists’ community memberships across time. Under low success bias, scientists frequently move between communities, showing there is open information flow and heterogeneous theory adoption. By the end of the simulation, all of the initially unaffiliated scientists have joined a community. In contrast, under high success bias, information flow becomes concentrated around a few highly successful scientists. Those with poorly performing theories are disproportionately likely to adopt ideas from this small subset, while scientists with moderately successful theories sometimes refrain from social learning altogether–their theories perform well enough to discourage abandoning them via social learning, but not well enough to merit being adopted by others. The resulting alluvial diagram (right) shows fewer, more stable communities with limited exchange between them.

Figure \ref{fig:alluv} panel B visualizes total information flow at the end of two representative simulations on a chord diagram. Each color denotes a distinct agent, with the ribbons showing the direction and magnitude of information exchange (e.g. a purple ribbon from purple to orange means orange adopted information from purple). Under low success bias, information circulates broadly and relatively symmetrically among all agents, reflecting a somewhat diverse and reciprocal exchange. However, under high success bias a few moderately successful scientists again cease participating in information exchange altogether, and most exchanges originate from the single most successful community (red), creating a significantly more centralized structure.

\subsection{Success Bias and Theory Development}\label{sec:SB and theory dev}
At first glance, success bias seems like an efficient way to ratchet up the overall quality of solutions in a community by allowing collective effort to concentrate on the most promising ones. Indeed, in our simulations, there was a significant positive correlation between success bias and the average perceived quality of theories produced (Pearson's $r(2299) = .17,  p < .001$, using log-temperature\footnote{Throughout, we report success bias as negative log-temperature , so that higher values correspond to stronger success bias. When correlating success bias with theory performance (measured as loss), we use positive log-temperature so that the sign of the correlation reflects the intuitive direction (positive = bias improves performance, negative = bias worsens it).}), which is created by prompting unsuccessful scientists to abandon their own theories (in part or in whole) in favor of those that appear more successful.

However, given the disconnect between perceived and actual theory success, it follows that optimizing for perceived success is not necessarily equivalent to optimizing for actual success of theories. We found that success bias was weakly negatively correlated with the community’s average actual theoretical performance ($r(2299) = -.052,  p =.012$), and was strongly correlated with a \emph{decline of the best} theories (the top-quartile actual performance across simulations) ($r(2299) = -.57,  p < .001$), and an \emph{improvement of the worst} (bottom quartile) ones ($r(2299) = .36,  p < .001$), thereby reducing the overall variance in theoretical success within the community (see Figure \ref{fig:illusions}, panel C). In other words, communities whose agents preferentially adopt theories based on perceived success avoid catastrophic failures at the expense of the possibility of developing truly outstanding theories.

Why might this be? We explore three possible explanations: 1) a reduction in the diversity of theories explored by the scientific community (section \ref{sec:SB and div}), 2) vulnerability to misleading signals of success (section \ref{sec:CrPC}), and 3) increased overfitting of theories (section \ref{sec:overfitting}). 

\subsubsection{Success bias reduces the diversity of developed theories}\label{sec:SB and div}
A substantial body of work suggests that epistemic diversity, the breadth of theories or solutions actively pursued by a community, plays a crucial role in collective problem-solving and discovery\cite{Campbell_2022, Hong_2004, Zollman_2010, Smaldino_2024, Page_2008, wu2023epistemic}. Diverse communities are more likely to explore a broader space of possibilities, avoiding premature convergence on local optima as opposed to global ones. So, our first explanation concerns the way success bias affects the diversity of theories explored by the agents. 

\begin{figure}
    \centering
    \includegraphics[width=1\linewidth]{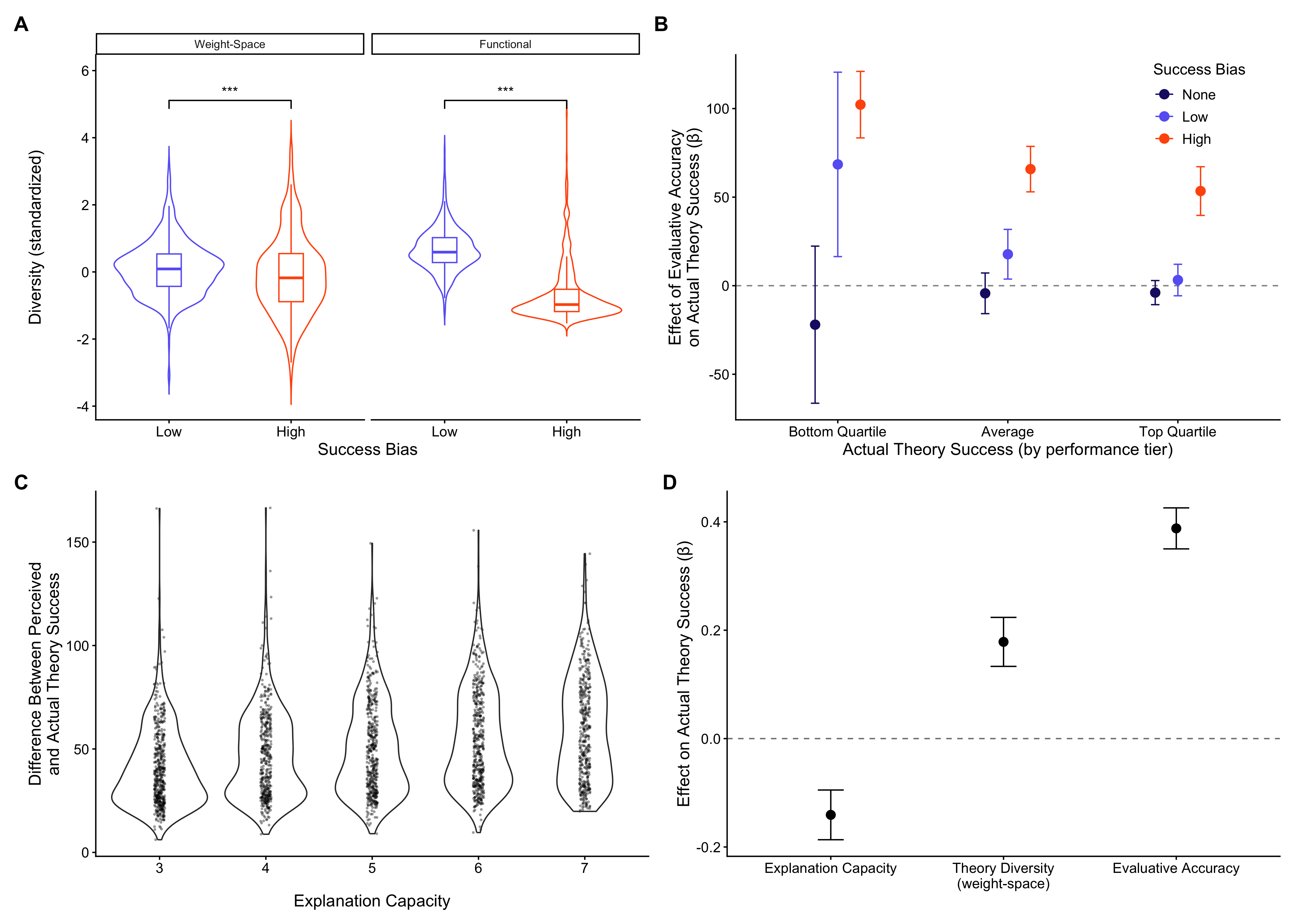}
    \caption{\textbf{Mechanisms underlying the effects of success bias.} \textbf{(A)} Theory diversity is significantly lower under high success bias than low for both measures of diversity, indicating that success bias homogenizes agents' theories. Note: both diversity measures are standardized (z-scored) independently to allow for comparison on a shared axis. \textbf{(B)} Regression coefficients and 95\% CI's for the effect of evaluative accuracy (CrPC) on actual theory success, split by performance tier and success bias regime. Under high success bias, evaluative accuracy \textit{substantially} improves bottom-quartile theories but has a diminishing effect on higher-performing ones. Under no success bias, evaluative accuracy has no significant effect at any performance tier. \textbf{(C)} Increasing explanation capacity widens the gap between perceived and actual theory success, indicating more overfitting: theories that explain more give agents a misleading sense of success. \textbf{(D)} Standardized coefficients from a multiple regression predicting actual theory success. Evaluative accuracy and theory diversity are associated with better outcomes, while explanation capacity is associated with worse outcomes.}
    \label{fig:explainers}
\end{figure}

High levels of success bias cause communities to converge on a narrow set of theories. When agents strongly prefer to adopt from apparently successful theories that outperform their own, they increasingly pursue the same favorable alternatives. We measure this convergence, and subsequent decrease in theory diversity, in two ways: in weight-space (comparing theories' internal parameters) and in function-space (comparing what theories actually predict about the phenomena each community studies, see section \ref{sec:theory eval}). Under high success bias, agents' theories become more similar in weight-space (median: high SB = 5.04, low SB = 5.24; Mann-Whitney $U=189,372,\;p<.001,\; n=576$ per group), and the predictions of these theories become more homogeneous (median: high SB = 5.45, low SB = 20.74; Mann-Whitney $U=241,028,\;p<.001,\; n_{\text{high}}=460,\; n_{\text{low}}=576$, see figure \ref{fig:explainers} panel A\footnote{Sample sizes for functional diversity are lower because some simulations fragment into many near-singleton communities where no community has enough members and evaluation points to compute within-community functional diversity.}).

The loss of diversity carries an epistemic cost---as agents within a community increasingly adopt the same apparently successful theories, they lose the variety of approaches required to effectively explore the space of possible new theories. This helps explain why success bias degrades the top quartile of theories produced, because without diverse exploration, communities are unlikely to stumble on theories that are genuinely better than those they already have, even if they are very good at discarding the worst ones.

\subsubsection{Evaluative accuracy filters out poor theories but cannot improve the best}\label{sec:CrPC}
Even though diversity fosters exploration, it alone is not sufficient for epistemic success. Diverse communities are more likely to generate high-quality theories by virtue of their exploration, but they do not always recognize or retain them. Simply having more candidate solutions is only helpful insofar as scientists possess \emph{evaluative accuracy}, or the ability to distinguish good theories from bad ones. We operationalize this idea using Centrality and Relative-Performance Correlation (CrPC), which measures how closely a theory’s popularity (Katz centrality of its owner) aligns with its \emph{actual} quality (relative to the mean across all other theories). Thus, a high CrPC implies that the community has high evaluative accuracy and successfully amplifies and shares actually good theories. Low CrPC, on the other hand, implies a divergence between perceived and actual quality, suggesting an illusion of understanding is at play.

Evaluative accuracy, however, remains latent unless agents act on their assessments of others' theories. Consequently, its actual epistemic impact depends on the degree of success bias. When a community has no success bias, agents don’t coordinate their imitation around perceived success, so the presence of misleading signals (low CrPC) has no impact on the community’s theory exchange. In high-bias regimes, however, the agents actively use these signals to guide adoption, so imitation is explicitly coordinated around perceived success, meaning evaluative failures can be amplified.

Figure \ref{fig:explainers} panel B shows this conditional dependence. We regressed actual theory success on evaluative accuracy (CrPC) separately for each tier of theory performance (bottom quartile, average, top quartile) and success bias condition. Under no success bias, evaluative accuracy has no relationship to theory outcomes at any performance tier (all CIs cross zero). But under high success bias, evaluative accuracy has a strong positive effect on average theory outcomes ($\beta = 65.8$, 95\% CI $[53.0, 78.7]$), meaning, in turn, that lower evaluative accuracy is associated with substantially lower average theory success. 

But even when communities \textit{can} distinguish good theories from bad ones, there's a functional asymmetry as to \textit{where} the benefits of this ability accrue. Evaluative accuracy has its largest effect on the worst theories ($\beta = 102.$, 95\% CI $[83.4, 121.]$), but almost half that effect on the best ($\beta = 53.4$, 95\% CI $[39.7, 67.2]$). Within the bottom quartile of theories, then, discernment acts like a filter---success-biased communities that possess high CrPC are able to aggressively reject poor theories by dropping them in favor of more successful ones. This reduces bottom-quartile error relative to unbiased communities (median error: high SB $= 114.3$, $n = 576$; no SB $= 168.0$, $n = 946$; Mann-Whitney $U = 119{,}845,\; p < .001$). In effect, evaluative ability and success bias work together to `raise the floor’ of theories produced. However, among the top-quartile of theories, even with a high evaluative ability, success-biased communities fail to outperform unbiased ones, even performing significantly worse (median top-quartile error: high SB $= 87.52$, $n = 144$, no SB $= 67.68$, $n = 946$; Mann-Whitney $U = 121{,}241,\; p < .001$). Although evaluative ability helps agents converge on what happen to be the best \textit{currently available} ideas, this convergence comes at the cost of the exploration needed to generate genuinely \textit{better} ones. Success bias raises the floor but actively lowers the ceiling.

\subsubsection{Success biased agents are misled by overfit theories}\label{sec:overfitting}
The reliability of a community's evaluative ability, as discussed above, is deeply entangled with overfitting. Overfitting occurs when a theory is excessively tailored to the idiosyncrasies of a subset of datapoints, giving it inflated accuracy on these data points but poor generalizability to new ones. In the context of neural networks, the risk of overfitting scales with the degrees of freedom a model has \cite{Williamson_2024}. In our simulations, we manipulate this by increasing the scientists’ explanatory capacity (the number of hidden neurons in their theories). While adding neurons gives a scientist the capacity to build more complex theories, it also provides the excess freedom to simply memorize random noise rather than learning the true underlying patterns. We find that increasing explanation capacity significantly degrades the community's evaluative accuracy (Spearman's $\rho(2299) = -.19, p < .001$). In fact, at the highest explanation capacity, evaluative accuracy is not significantly different from zero (i.e. agents adopt from others non-randomly, but their selection on the basis of success is no better than random) (Wilcoxon signed-rank $(N=439)$, $V = 50431, p = .421, \text{median} = .003$). This degradation of evaluative ability is consistent with the illusion of understanding because explanation capacity also significantly widens the gap between perceived and actual theory success (Spearman's $\rho(2299) = .28, p < .001$) (see Figure \ref{fig:explainers}, panel C). While this effect is universal, its consequences follow the asymmetry established above whereby unbiased communities ignore the noisy signal overfitting creates, but success-biased ones actively coordinate around it.

\subsection{Optimizing for Perceived Success}\label{sec:opt}
While the previous results outline the epistemic cost of success bias, they leave open the question of strategic adaptation. If scientists are rational agents motivated to maximize their theoretical success, which social learning strategies would they naturally adopt? To answer this, we conducted an inverse design experiment using Bayesian optimization \cite{Shahriari_2016}. We treated the social parameters of the model (success bias, community bias, degree of sociality and explanation capacity) as tunable variables in order to find the configuration that \textit{maximizes} the perceived success of scientists. We repeated this procedure across six environments varying in problem complexity and density, and used two phases of optimization: an initial Sobol sampling phase to explore the parameter space more broadly \cite{Sobol_1967}, then a Gaussian process-guided Bayesian optimization to refine the parameters towards an optimal configuration.

\begin{figure}
    \centering
    \includegraphics[width=1\linewidth]{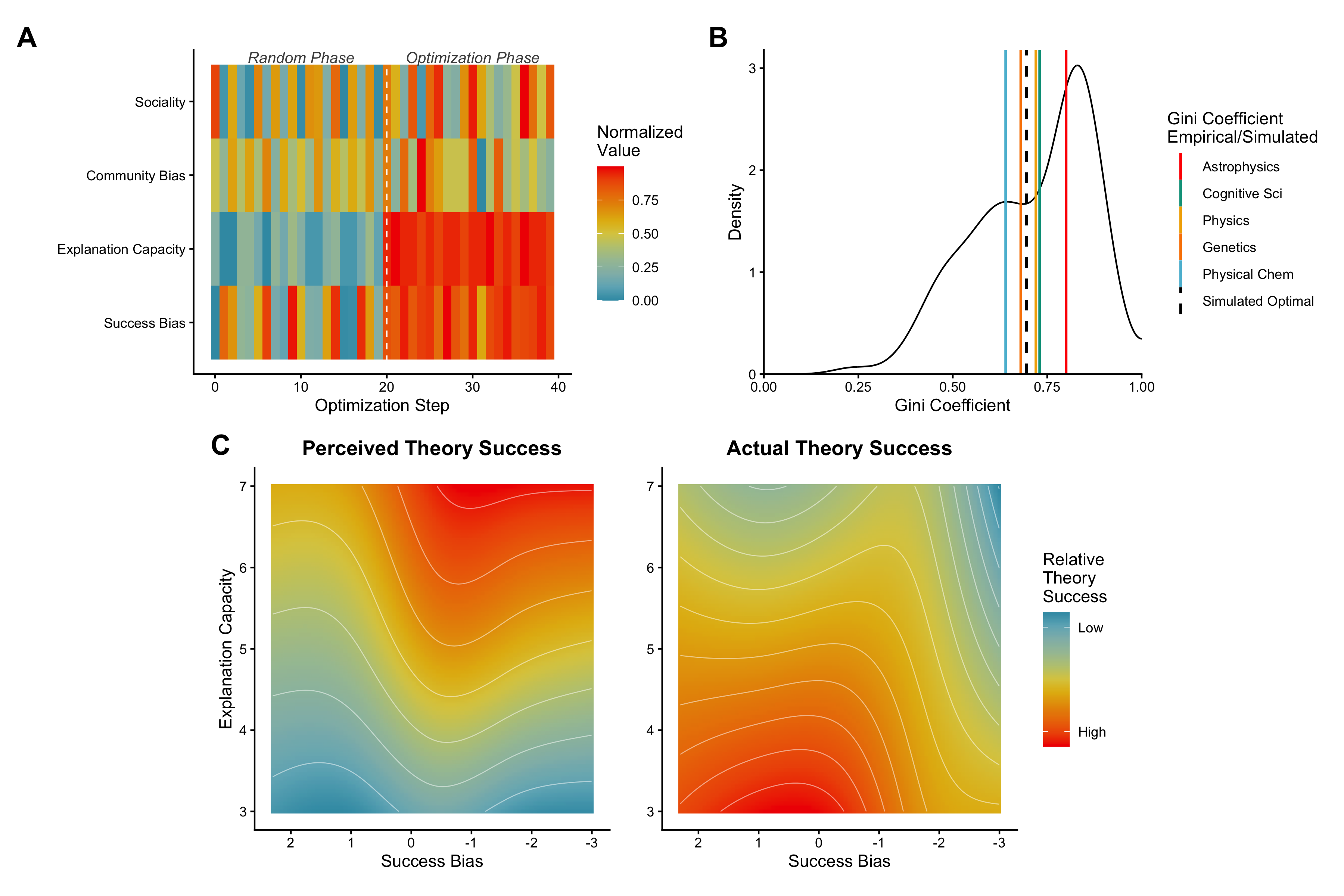}
    \caption{\textbf{Results of Optimization.} \textbf{(A)} Parameter values explored during random (Sobol) and optimization (GP-guided) phases, averaged across six environments. The optimizer converges strongly on high success bias and maximum explanation capacity, while remaining agnostic to community bias and sociality. \textbf{(B)} Distribution of Gini coefficients during the optimization phase. Vertical colored lines show empirical citation inequality; the dashed line marks the mean optimal Gini (0.69), which sits squarely in the range of those empirically observed. \textbf{(C)} GAM-predicted perceived (left) and actual (right) theory success across configurations of success bias and explanation capacity. The regions that maximize perceived success (high bias, high capacity) correspond to average or below-average actual performance, illustrating the divergence between perceived and actual theoretical quality.}
    \label{fig:opt}
\end{figure}

We found that the optimizer had a strong and consistent preference for two of the four parameters, namely success bias and explanation capacity. Success bias converged toward high values across all 6 environments (GP-phase median temperature = $.67$ ($n=120$), Sobol-phase median $= .14$ ($n=120$); Mann-Whitney $U = 10,504,\; p < .001$), with variance decreasing by 75\%. Explanation capacity converged even more decisively reaching its maximum value (8) in every environment’s optimization phase (GP-phase median $= 8.0$, Sobol-phase median $= 5.5$ ; Mann-Whitney $U = 12,668,\;p < .001$), with variance decreasing by 95\%. In contrast, community bias didn’t show significant convergence (Mann-Whitney $U = 7,557,\;p = .504$), and sociality showed only nominal convergence (Mann-Whitney $U=8,549,\;p=.012$), suggesting that these parameters’ roles were overshadowed by the influence of success bias and explanation capacity (see Figure \ref{fig:opt}, panel A).

The convergence on high success bias is consistent with the results outlined in section \ref{sec:SB and theory dev} where we found success bias to be a reliable predictor of the perceived success of theories developed. That said, the strong preference for high explanation capacity is more revealing. As discussed in section \ref{sec:overfitting}, high explanation capacity increases the ability of agents to overfit their theories, which inflates perceived success. The optimizer seems to exploit precisely this vulnerability; in combining high success bias as well as high explanation capacity, agents reach configurations that seem maximally successful. However, as outlined in the results above, they may be deceptively so. Across all six environments, configurations selected by the optimizer near the end of optimization (last 4 steps) sat at the 81st percentile (95\% bootstrap CI $[72, 87]$) in perceived success but only the 32nd percentile (95\% bootstrap CI $[15, 43]$) in actual success relative to randomly sampled configurations. Thus performing worse than the typical random configuration at the actual task while \textit{appearing} better than most.

\subsubsection{Optimizing for perceived success reproduces real-world inequality}
These results describe what rational success-maximizing agents in our model would do, but do real scientific communities exhibit the same social structures that such optimization produces? Although we can’t measure the degree to which working scientists are tuned to theoretical success, success bias generates a measurable downstream effect of inequality in epistemic influence. When agents preferentially adopt from the most apparently successful others, influence concentrates around these handful of individuals instead of being distributed more broadly across the scientific community. In our model, we quantify this inequality using a Gini coefficient over theory adoptions (cumulative adoptions against cumulative agents), where if the Gini coefficient of a simulation is high, a small number of agents’ theories have an outsized influence on the community at large. Importantly, Gini emerges as property of the simulation, rather than being set by any particular parameter; the above social parameters all contribute in shaping it.

To situate our model’s optimized social parameters against real-world science, we estimated inequality in knowledge-making across several fields using citation data from Scopus. We use citations to compute a proxy of information exchange in science, since citations track, albeit at a coarse grain, whose work is being attended to and built upon. Here, again, a high Gini coefficient implies that a small number of authors hold outsized influence on the literature. For each field, we randomly sampled 20,000 articles written between 2000 and 2020 and computed its Gini coefficient (see Section \ref{sec:gini_methods}). Across fields, we found Gini coefficients ranging from $0.63$ to $0.80$: Astrophysics ($0.80$, 95\% CI $[0.79, 0.80]$), Cognitive Science ($0.73$, $[0.72, 0.73]$), Physics ($0.72$, $[0.72, 0.73]$), Physical Chemistry ($0.64$, $[0.63, 0.64]$), and Genetics ($0.63$, $[0.62, 0.64]$). While these estimates come with caveats---there are many possible causes of an unequal citations distribution that go beyond success bias: e.g., citations simply favor older publications, fields vary in size and citation norms, and citations are not always positive references---we note that the Gini coefficient that emerged from our optimization procedure ($0.69$, 95\% CI $[0.62,  0.76]$) sits squarely within the range of those observed in the real world (see Figure \ref{fig:opt}, panel B), in this analysis as well as others \cite{Nielsen_2021}. This suggests that the levels of inequality we observe in the real world might arise, at least in part, from scientists individually optimizing for perceived success. But as our earlier results indicate, what \textit{seems} optimal and what \textit{is} optimal may be misaligned: the social configurations that maximize perceived success do not maximize, and in fact often actively undermine, actual scientific progress (Figure \ref{fig:opt}, panel C).

\section{Discussion}
Social learning lies at the heart of the scientific project. The cumulative growth of knowledge, which is arguably science’s greatest achievement, relies on scientists’ capacity to learn from one another, build upon existing frameworks, and selectively inherit and iterate on the ideas of one’s predecessors \cite{Boyd_1988}. This introduces a critical question: \textit{how} should scientists learn from one another? Among the many possible strategies, success-biased learning offers what appears to be a rational solution \cite{Laland_2004}. By preferentially adopting the ideas that seem most successful, communities can, in principle, efficiently concentrate their collective efforts on the most promising solutions without expending effort on dead ends \cite{Rendell_2010}. This strategy also aligns with intuitive notions of optimization, wherein consistently following the gradient towards the most successful solutions will lead to local if not global optima.

Our work, as well as that of others \cite{Wu_2024}, presents a very different view of success-biased learning in which its apparent logic is undermined by the illusion of understanding. More specifically, when the perceived and actual success of theories is misaligned, success bias coordinates imitation around the wrong targets. As a consequence, the diversity of ideas pursued reduces, and scientists converge prematurely on mediocre solutions, discarding ideas that underperformed early but might have matured into something excellent.

By modeling science with a multi-agent model, we found that irrespective of the social learning strategies employed, agents systematically overestimate the quality of their theories compared to the theories' actual performance on the ground truth. This gap grows as the complexity of the phenomenon under study increases. One reason for this illusion is overfitting: the more expressive a scientist’s theoretical framework, the easier it becomes to fit the idiosyncrasies of limited data rather than capturing some of the underlying regularities, if they exist. These overfit theories register as highly successful, predicting their own data well, but capture little of the ground truth. Success bias then compounds this as overfit theories attract imitation and the community converges around them. The more complex the problem, the less scientists can trust their own assessments, and by extension, the less they should trust the apparent success of others’ work.

Success bias fails when the community using it can’t accurately identify quality, but even when it can, success bias primarily benefits the worst theories rather than the best. Success bias helps poor performers catch up by encouraging them to imitate better theories, but offers nothing to those already at the frontier. Meanwhile, by amplifying convergence on currently successful approaches, success bias reduces the diversity of solutions that might otherwise have generated new breakthroughs.

These dynamics are not solely theoretical. Success bias pervades the actual practice of science through many means: graduate students choose advisors and labs often on the basis of the success of the theories they champion, hiring committees favor successful candidates, and funding bodies commonly allocate resources based on the anticipated success of a given project or else on metrics of past success \cite{Merton_1968, Jimenez_2019, Clauset_2015, Fanelli_2010}. Each of these mechanisms direct attention, and thus the future trajectory of inquiry, toward what has already appeared to work well. It’s worth noting that this directing of attention is exacerbated in the real world by other biases, such as preferences based on prestige, title, seniority, and other indicators which serve as an even noisier proxy for scientists' success. Our model, in this respect, is quite conservative. We simulate scientists who learn from others solely based on the performance of their theories, rather than using more obviously problematic strategies like deference to seniority, institutional affiliation, or popularity. That success bias proves to be damaging even in these relatively meritocratic and epistemically \textit{virtuous} conditions suggests that the problem runs deeper than poor heuristics or irrationality.

In fact, if we presume that scientists rationally configure their social learning behaviors in order to optimize for the perceived success of their own theories, we have seen (in section \ref{sec:opt}) that they may do so to the detriment of actual epistemic progress. In our model, the configuration of social behaviors that optimizes perceived theoretical success among agents is precisely the kind of configuration that undermines its actual counterpart (see Figure \ref{fig:opt}, panel C, and section \ref{sec:opt}). Further, the kind of knowledge making inequality that `optimal’ success bias induces in the model is very close to what we observe in real world citation networks, making it possible that we are indeed subject to illusions of understanding that arise out of our deference to apparent success.

\subsection{The Generality of Success Bias \& the Selection Problem}
Although our simulations were designed to reflect key features of scientific practice, the underlying mechanisms driving success bias’ detrimental effects seem likely to appear in any community of social learners engaging with sufficiently complex problems. The core mechanism does not depend specifically on features of scientific learning, but rather emerges from the structure of the learning problem itself, the kinds of noisy success signals it might produce, and their social amplification. The domains in which success bias is likely to have detrimental effects share a common feature: they are what some call \textit{strong-link} problems, where outcomes depend disproportionately on the quality of the \textit{best} solutions, rather than eliminating the worst \cite{Mastroianni_2025}. If at all, success bias seems better suited to \textit{weak-link} problems, where avoiding failures matters more than achieving excellent best solutions. 

This failure mode extends far beyond science. Success bias belongs to a broader class of errors in which optimizing (i.e., selecting) for the wrong (or noisy) criteria can undermine the process of search it's meant to guide. The fundamental difficulty is not only that success metrics are noisy proxies for epistemic quality---though they are---but that in \textit{genuinely open-ended search}, we have no reliable way of knowing what good solutions will look like in advance, and thus no way of selecting among possible intermediary steps towards such solutions. The history of science provides ample evidence for this point: theories that were once considered cutting-edge appear, to the well-informed scientist of today (and often the average person), not merely incomplete but deeply misguided, or even absurd. This is not a flaw that better metrics or evaluation can in principle fix. When the destination is unknown, as it is in science and many other domains, any fixed notion of ``progress toward it" is liable to mislead \cite{Stanley_2015, Goodhart_1975}. Success bias may amplify this problem by concentrating collective effort around whatever currently appears successful.

But even if success signals were reliable, success bias would \textit{still} pose the problem of over-exploitation. Converging on what already works necessarily entails leaving by the wayside alternatives, and reducing the degree to which scientists explore the possibilities available to them \cite{March_1991, Chang_2017, Chang_2007, Wimsatt_2007, Massimi_2022}. In environments where the best theories are not immediately evident, a category to which science surely belongs, exploiting existing solutions, over the long term, might prove to be a poor tactic for finding innovative new ones. 

Indeed, from the perspective of active inference, a framework in which agents balance exploiting what they believe works with exploring to reduce their own uncertainty \cite{Friston_2014}, this over-exploitation is not merely suboptimal but formally irrational. There, the rational balance between exploitation and exploration is governed by the precision of an agent's evaluative signals: when an agent can reliably distinguish good theories from bad ones, it should exploit; when it cannot, exploration becomes the Bayes-optimal response, because sampling unfamiliar territory is the fastest way to gain information that can actually reduce uncertainty. This might explain why agents in our model, when optimizing their social behavior toward perceived success, converge so strongly on success-biased strategies and thereby reduce actual theory success. These agents optimize as though their evaluative signals were trustworthy, without recognizing that those signals are systematically noisy. Had they noticed this (i.e., had they modeled the gap between perceived and actual theory quality) the optimal strategy could shift toward exploration.\footnote{Of course, this raises the far deeper difficulty that accurately estimating the reliability of one’s own evaluations might be intractable from within the system. Scientists cannot easily step outside their own evidence to assess how well their proxies for quality track actual quality, which is precisely the epistemic constraint that gives rise to illusions of understanding in the first place.}  A rational agent that accurately estimated the unreliability of its own assessments, would be less success-biased, not more---but achieving that self-knowledge seems to be the hardest part of the problem.

Future work should consider other factors that might counteract this over-exploitation in science; for instance, competition for scarce rewards such as faculty positions and elite publications may motivate scientists to pursue riskier ideas associated with higher rewards \cite{gross2025competition}.

\section{Recommendations and Limitations}

If reliable evaluation of theories is indeed difficult, then what follows for how science should be organized? One possibility would be to recognize that making decisions based on apparent success can inadvertently weed out excellent ideas, so maintaining a diverse set of theories should serve as a hedge against our own evaluative limitations. Science would then have to be structured to tolerate disagreement and protect minority approaches \cite{Goldstone_2013}. Depending on the constraints and social organization of a field, epistemic progress may be better served by adopting a pluralist approach, and developing parallel systems of understanding, even if they are mutually incommensurable \cite{Chang_2017, Wimsatt_2007, Feyerabend_1993, Massimi_2022}.

An obvious objection is that unbounded pluralism is deeply impractical. The space of possible theories is incredibly vast, and without some mechanism for directing collective attention, science risks diffusing its efforts too thinly to make progress on any front. This is precisely why selectionist accounts of science are appealing: they promise a principled mechanism, analogous to natural selection, for narrowing the space of theories down to those worth pursuing \cite{Campbell_1960, Popper_1979, Hull_1990}. Our results do not dispute the need for selection, but they dispute the form it currently takes. Success bias, as modeled here, selects on specific theories---agents adopt particular theoretical structures wholesale, not classes of theories that happen to predict well. This is a highly restrictive form of selection, and our results suggest it carries a corresponding cost in diversity and long-term exploratory outcomes. A less restrictive alternative might involve selecting among \textit{families} of theories that explain the same phenomena, preserving internal variation that could enable a scientific community to remain exploratory and to have the capacity to respond more effectively when surprising new evidence arrives \cite{Chang_2012}.

The nature of our model, however, limits its application to the specifics of scientific practice. As our results show, the outcomes of different social learning strategies depend heavily on the structure of the ground truth and scientists’ theory-making capacities. Our results, then, may have little application to the real world where neither of these things is known with precision. Furthermore, in each simulation, the social learning strategy was a global parameter: agents were always homogeneous within a given simulation run. However, it could be that a diversity of approaches in a population of learners produces very different results, and further work will have to explore mixtures of learning styles that produce optimal outcomes with respect to both perceived and actual theoretical success. Another limitation concerns the ground truth which was represented as a series of n-dimensional bounded Gaussians surrounded by random noise. Future work could explore whether or not the results presented here generalize across different ground truths generated by the same mechanism, and for qualitatively different structures altogether. 

Other limitations in our model include idealization of agents' epistemic actions: agents in our model possess purely statistical abilities (building a model that predicts output values from inputs), which fails to capture the diverse modes of reasoning actual scientists employ; our agents also learn via backpropagation and are only capable of possessing a single theory at any given time. In reality, scientists learn in a variety of ways and sometimes work on multiple theories simultaneously \cite{Legare_2018, Chang_2014}, adopting elements from a dominant paradigm while continuing to develop their own ideas in parallel, or setting some ideas aside to return to them later. 

\section{Conclusion}
Science is a fundamentally social enterprise, and some of its most important dynamics---which ideas spread and are built upon---are directly shaped by the ways in which scientists choose to learn from one another. Here, we found that using the perceived success of a theory as signal for social learning might actively hinder collective discovery. Success-biased learning reduces the diversity of theories explored, and rewards overfitting: theories that explain collected data well but fail to generalize. Even when communities \textit{can} successfully identify which theories are genuinely best, success bias only raises the floor by filtering out the worst theories, but fails to raise the ceiling by supporting scientists to discover new, even better accounts of the world. Worse, when agents optimize their own social behavior to maximize the perceived success of their theories, they paradoxically undermine their actual performance.

These findings suggest that our ability to evaluate new theories is far more limited than is often assumed. We have built institutions that reward optimization, treating success as a reliable compass and selection as a virtue. But if apparent success is as likely to reflect the limits of our evidence as the quality of our ideas, then progress may depend less on selecting the right theory and more on maintaining a plurality of explanations worthy of a world that is richer, stranger, and more imperfect than any single theory can capture.

\newpage

\newpage
\bibliography{cites_0} 
\bibliographystyle{sciencemag}

\newpage

%%%%%%%%%%%%%%%% ACKNOWLEDGEMENTS %%%%%%%%%%%%%%%

\section*{Acknowledgments}

%%% FOR SUBMISSION VERSION YOU CAN REPLACE WITH "REDACTED FOR PEER REVIEW" all sections except the last two sections which should be left unchanged

The authors thank Mirta Galesic for productive discussions that led to the improvement of this work.

\paragraph*{Funding:}

A.~L. was supported by the National Science Foundation under Award No.\ 2349052, the The Bengier Foundation grant to SFI, and the GEAR grant as part of the Symbolic Systems Department at Stanford University. 
M.~D. was supported by the SFI Lou Schuyler award.

\paragraph*{Author contributions:}

\textbf{A.L.} Conceptualization, Methodology, Software, Validation, Formal Analysis, Investigation, Data Curation, Writing -- Original Draft, Writing -- Review \& Editing, Visualization. \textbf{M.D.} Conceptualization, Methodology, Formal Analysis, Investigation, Writing -- Original Draft, Writing -- Review \& Editing, Supervision.

\paragraph*{Competing interests:}
There are no competing interests to declare.
\paragraph*{Data and materials availability:}
All code, data, and simulation outputs are available at \url{https://osf.io/u3xc9/?view_only=309cb49465554a869932147b1ab74b50}.

\newpage

%%%%%%%%%%%%%%%% SUPPLEMENTARY TEXT %%%%%%%%%%%%%%%

\subsection*{Supplementary Text}

\subsubsection*{Robustness of success bias comparisons to other split criteria}

\begin{figure} % Do not use \begin{figure*}
	\centering
	\includegraphics[width=1.0\textwidth]{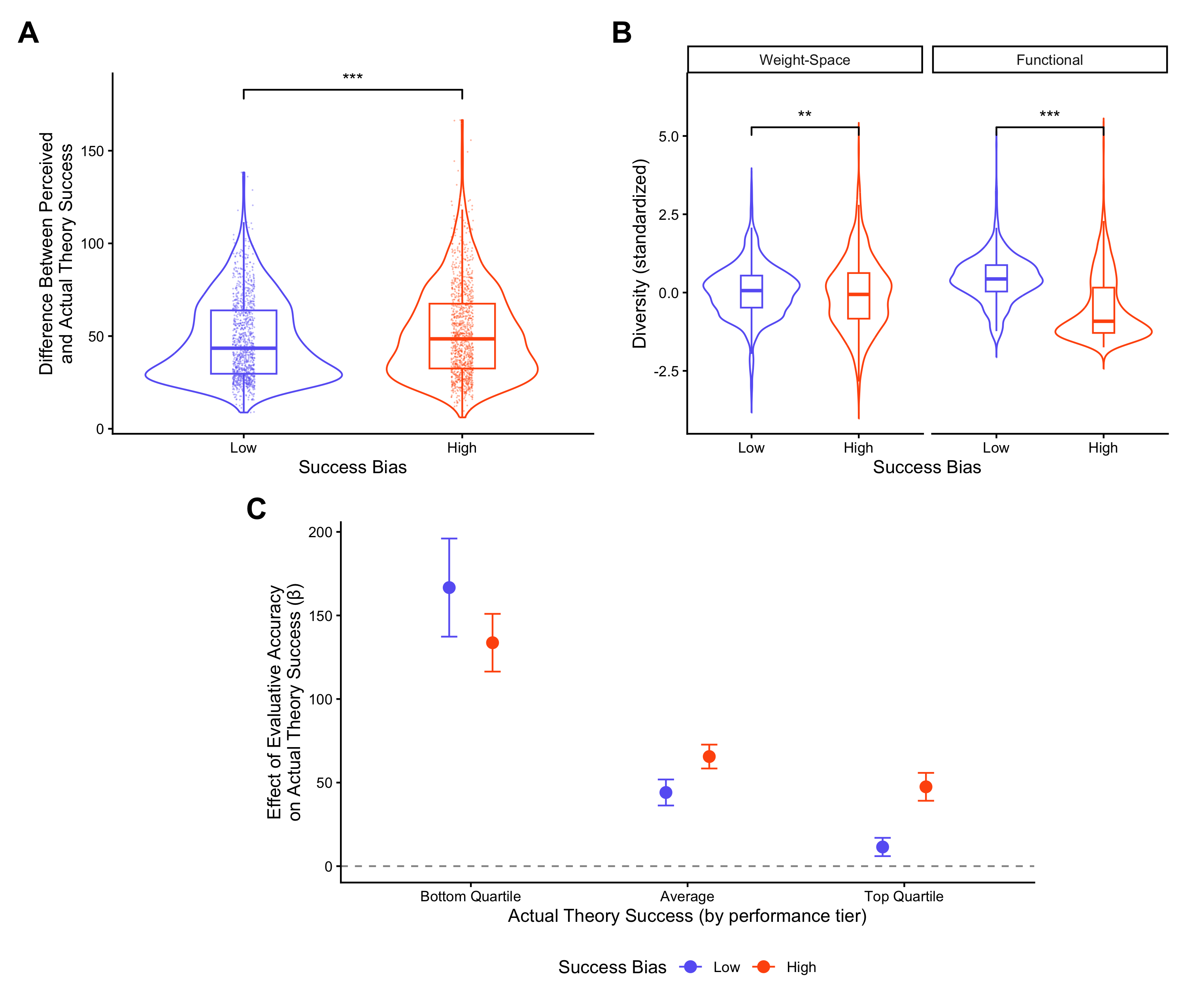}

	% Captions go below figures
	\caption{\textbf{Comparisons of high-versus-low success bias using a median split.}
		Same as in Figures~\ref{fig:illusions} and \ref{fig:explainers}, but for the median split. (\textbf{A}) The gap between perceived and actual success is significantly higher under high success bias as split by the median (Mann-Whitney $U = 603,959,\; p < .001$, $n= 1150$ per group), (\textbf{B}) Under high success bias by a median split, weight-space diversity decreases similarly (median: high SB = 5.16, low SB = 5.25; Mann-Whitney $U=711,502,\;p=.002,\; n_{\text{high}}=1150,\; n_{\text{low}}=1151$ per group), and functional diversity still drops significantly (median: high SB = 7.96, low SB = 20.87; Mann-Whitney $U=898,575,\;p<.001,\; n_{\text{high}}=999,\; n_{\text{low}}=1149$, (\textbf{C}) Unlike the quartile split, the effect of CrPC on the bottom quartile of theory performance is not significant under the median split (interaction $p = .053$). The effects at the average and top quartile are still significant (both $p < .001$).}
	\label{fig:sup_median} 
\end{figure}

Throughout the main text, we compare high and low success bias conditions 
using a quartile split, keeping the top and bottom 25\% of simulations 
(by temperature) and removing the middle 50\%. This creates some contrast 
between conditions but is necessarily an arbitrary threshold. To ensure 
our results are not just an artifact of this choice, we replicate all 
high-versus-low comparisons using a median split, which keeps all 
simulations and assigns them to whichever half of the temperature 
distribution they fall in (Supplementary Figure \ref{fig:sup_median}).

Results are qualitatively identical across both approaches, with one 
minor exception-- under the median split, the effect of evaluative 
accuracy (CrPC) on bottom-quartile theory success no longer differs 
significantly between high and low success bias conditions 
(Supplementary Figure \ref{fig:sup_median}, panel C). This is because the median split includes simulations with moderate levels of success bias in the ``low" group, which dilutes the contrast. 

% If your supplement is very short you might need to uncomment the following line to avoid
% layout problems with the figures and tables.
\newpage

%%%%%%%%%%%%%%%% SUPPLEMENTARY FIGURES %%%%%%%%%%%%%%%

\end{document}